\documentclass[aps,prd,nofootinbib,showpacs,twocolumn,groupedaddress,amsmath,amssymb,preprintnumbers]{revtex4}

\usepackage{graphicx}
\usepackage{dcolumn}
\usepackage{bm}
\usepackage{hyperref}
\usepackage{color}
\usepackage[pdftex]{epsfig}  
\usepackage[normalem]{ulem}
\usepackage{subfig}

\def\square{\kern1pt\vbox{\hrule height 1.2pt\hbox{\vrule width 1.2pt\hskip 3pt
   \vbox{\vskip 6pt}\hskip 3pt\vrule width 0.6pt}\hrule height 0.6pt}\kern1pt}

\newcommand{\be}{\begin{equation}}
\newcommand{\ee}{\end{equation}}
\newcommand{\bea}{\begin{eqnarray}}
\newcommand{\eea}{\end{eqnarray}}
\newcommand{\nn}{\nonumber}

\newcommand{\vv}[1]{\mathbf{#1}}
\newcommand{\kv}{\mathbf{k}}
\newcommand{\xv}{\mathbf{x}}

\newcommand{\fnl}{f_{\rm NL}}

\def\dthreee#1#2{\frac{d^3{#1_{#2}}}{(2\pi)^3}}

\begin{document}
\preprint{IGC-16/5-1}
\title{On the space of non-Gaussian fields with single-clock bispectra}

\author{Bekir Baytas}
\email{bub188@psu.edu}
\author{Sarah Shandera}
\email{shandera@gravity.psu.edu}

\affiliation{Institute for Gravitation and the Cosmos, 
The Pennsylvania State University, University Park, PA 16802, USA}

\begin{abstract}

We develop a mathematical construction of non-Gaussian fields whose bispectra satisfy the single-clock inflation consistency relation. At the same order that our basis for bispectra recovers the two simplest single clock templates, we also find a third orthogonal template which has the single clock squeezed limit, peaks in folded configurations, and has very small coupling in the equilateral limit. We explore the map between templates and operators in a very general Lagrangian for single-clock fluctuations and find no significant overlap between the new template and models in the literature. We comment on the physical implications of this conclusion. Our findings add support for the idea that both theory and data driven considerations will be best served if next generation non-Gaussianity constraints are made in a basis that uses the degree of coupling between long and short wavelength modes as an organizing principle.  

\end{abstract}


\maketitle

\section{Introduction}
\label{sec:intro}
The expectation for interesting threshold levels of non-Gaussianity comes from our current understanding of the space of likely models for inflation and the primordial fluctuations. The {\it Planck} satellite bounds demonstrate that the fluctuations are very close to Gaussian \cite{Ade:2015nlj}, but the constraints have not crossed the theoretically interesting threshold to rule out entire classes of physics beyond single field slow-roll \cite{Ade:2015ava}. Future constraints on non-Gaussianity will continue to test various inflation scenarios but can also have implications for our understanding of the universe that are independent of the origin of the fluctuations. A particular part of the non-Gaussian space, the coupling between long and short wavelength fluctuations, is of interest from both of these viewpoints. Purely from a statistical perspective, coupling between long and short wavelength modes introduces additional cosmic variance in the map between observations and theory \cite{Bartolo:2012sd,Nelson:2012sb,Nurmi:2013xv, LoVerde:2013xka,Thorsrud:2013mma, LoVerde:2013dgp, Thorsrud:2013kya, Byrnes:2013ysj, Bramante:2013moa,Baytas:2015nja,Adhikari:2015yya,Bonga:2015urq}. We call this super cosmic variance since it comes from non-Gaussian coupling of the observed modes to the unknowable super-horizon modes. From the inflationary theory perspective, the presence or absence of this coupling is a tool to qualitatively distinguish the number of degrees of freedom relevant for generating the inflationary background and fluctuations: In so-called `single-clock' models, short and long wavelengths are decoupled, so that fluctuations in any post-inflationary patch have intrinsic correlation functions\footnote{The intrinsic correlations may be different from correlations an observer at a particular point sees, due to projection effects \cite{Dai:2015rda,Dai:2015jaa}.} entirely determined by the background at the moment the fluctuations exited the horizon \cite{Senatore:2012wy,Joyce:2014aqa,Mirbabayi:2014zpa,Assassi:2012zq,Goldberger:2013rsa,Cheung:2007sv,Creminelli:2004yq,Young:2014oea}. The amplitude of long wavelength fluctuations is irrelevant. But, if fluctuations of a light degree of freedom other than the `clock' field contribute to the observed curvature perturbations, long-short mode coupling generically occurs \cite{Linde:1996gt, Moroi:2001ct, Lyth2001, Enqvist:2001zp, Dvali:2003em,Zaldarriaga:2003my,Chen:2009zp, Chen2010a,LoVerde:2013dgp,Bonga:2015urq, Kenton:2016abp,Kenton:2015lxa}. Interesting coupling between long and short wavelength modes can occur at any order in correlation functions, beginning with the bispectrum (Fourier space 3pt function). The consistency relations for single clock inflation require, among other things, that the bispectrum in the squeezed limit ($k_l\equiv k_1\ll k_2\approx k_3$) depend no more strongly on the long wavelength mode than $1/k_l$. Requiring such a scaling in the ``initial conditions" for the hot big bang universe (even without a dynamical model for generating it) ensures that these bispectra lead to no additional cosmic variance uncertainties in the power spectrum of the perturbations.

The importance of long-short mode coupling, both as a discriminator of inflationary scenarios and as a source of cosmic variance relating observations to theory, suggests that it be used to organize model-independent tests of non-Gaussianity. Such an approach was developed for the bispectrum by Byun and Bean in \cite{Byun:2013jba}, who demonstrated that such a basis efficiently covers a wide range of inflationary models in the literature. (There are other basis proposals, either overlapping or complementary to this one, and organized by other considerations \cite{Fergusson:2008ra, Meerburg:2010ks,Byun:2015rda}). They also found a basis template with the single-clock degree of long-short mode coupling but that was not well covered by the two standard single-clock templates. Since that work, the space of templates from single-clock inflationary Lagrangians has continued to grow, eg \cite{Behbahani:2014upa}, and the implications of cosmic variance consequences of mode-coupling have been developed \cite{Thorsrud:2013mma, LoVerde:2013dgp, Thorsrud:2013kya, Byrnes:2013ysj, Bramante:2013moa,Baytas:2015nja,Adhikari:2015yya,Bonga:2015urq}. In addition, the {\it Planck} satellite's tightest constraints on non-Gaussianity were published \cite{Ade:2015nlj}, and the focus for the next generation of constraints on single-clock scenarios has turned to Large Scale Structure surveys \cite{Byun:2014cea, Raccanelli:2015oma}.

Looking toward the future of non-Gaussianity constraints, in this paper we revisit the relationship between elements in a mathematical basis for non-Gaussian fields and the set produced by inflation. We are particularly motivated by asking whether or not a detection of a shape satisfying the single-clock consistency relation rules out multi-source scenarios. At one level, certainly not: knowing the form of one correlation function in a non-Gaussian field implies nothing about the shapes of the other correlation functions. In particular, there exist trispectra that generate equilateral shape bispectra in biased sub-volumes through long-short mode-coupling \cite{Baytas:2015nja}. These cannot be single-clock models, but they would be entirely consistent with the detection of an equilateral bispectrum\footnote{Such an effect may be detectable through constraints on higher order correlations, but requires sufficiently precise constraints of more general correlations, which will be challenging.}. However, even without cosmic variance from higher order correlations, we can ask whether every bispectrum that has the single-clock scaling in the squeezed limit can be generated (naturally or not) by single-clock inflation. The templates for single-clock inflation contain additional signatures of inflationary physics in momentum configurations away from the squeezed limit, so by constructing a basis and focusing on templates so far not in the literature we might hope to either ``reverse engineer" interesting single-clock inflation scenarios or find some additional physical limits on what they can produce. 

To address the relationship between the space of bispectra with no super cosmic variance and single-clock inflation models, we begin in the next section by generating non-Gaussian fields as non-linear, non-local functionals of a Gaussian field. The field will be fully specified (so all correlation functions can be computed and the field can be numerically generated), and one can add new terms order by order to control the details of higher order correlations. We focus on the first non-linear term, which is quadratic in the Gaussian field and so gives a tree-level bispectrum. We generate the space of fields by characterizing the quadratic term according to the number of inverse derivatives allowed per field, and then restrict the set by requiring the squeezed limit of the bispectrum diverge no more strongly than $1/k_l$. We call this set the super cosmic variance free set since there is no significant long-short mode coupling.  At the same order that this procedure recovers the two simplest single clock templates, we also find a third, orthogonal bispectrum that peaks in folded configurations and has very small coupling in the equilateral limit. We map this expansion to the bispectral basis of the Byun and Bean and find that our set of templates at that order can be expressed as a linear combination of theirs. We then examine in detail the relationship of the additional, ``folded-only" template to a very general effective Lagrangian for single-clock fluctuations (Section \ref{sec_singleclock}). We find that the folded-only template is still not found in the single-clock literature and discuss the reasons for small overlap in various cases. We conclude in Section \ref{sec_discuss} with some discussion of the implications of these results and how they can be extended.

\section{Generating super cosmic variance free models}
\label{sec_generating}

In this section, we first review a procedure to generate an expression for a non-Gaussian field with a bispectrum that matches the equilateral template. Then we show how the procedure may be generalized to non-Gaussian fields with the other standard (``orthogonal") template proposed for single-clock inflation. In the process we will uncover an additional bispectral template that also has the single-clock squeezed limit.

\subsection{Constructing a non-Gaussian field with an equilateral bispectrum}
\label{section_equil}
The lowest order statistic that is zero for a Gaussian field but non-zero more generally is the three-point function. Assuming the non-Gaussian field $\Phi$ is homogeneous, the bispectrum $B(\kv_1,\kv_2,\kv_3)$ is defined by
\begin{align}
\label{eq:threepoint}
\langle\Phi(\kv_1)\Phi(\kv_2)\Phi(\kv_3)\rangle=(2\pi)^3\delta^3(\kv_1+\kv_2+\kv_3)\;B(\kv_1,\kv_2,\kv_3)\;.
\end{align}
We also restrict to isotropic bispectra. In a generic single-clock inflation model, the form of the bispectrum need not be particularly simple. However, the bispectra of many models can be well approximated by  ``templates" built from a sum of terms that are each simple products of the three momenta \cite{Senatore:2009gt, Byun:2013jba, Fergusson:2008ra}. For example, the equilateral template is:
\bea
\label{eq:scaleinv_equil}
B_{\rm equil}(k_1, k_2, k_3) = 6 \fnl^{equil} A_{\Phi}^2\bigg[-\left(\frac{1}{k_1^3k_2^3} +{\rm 2\;perm.}\right) \nn \\
-\frac{2}{(k_1k_2k_3)^2}+\left(\frac{1}{k_1k_2^2k_3^3}+{\rm 5\;perm.}\right)\bigg]\;.
\eea
We have assumed a scale-invariant power spectrum for simplicity, with amplitude $A_{\Phi}^2=k^3P_{\Phi}(k)$, and $f^{equil}_{NL}$ is the amplitude parameter of the bispectrum, defined at the equilateral point in momentum space ($k_1=k_2=k_3$). Notice that even though the individual terms in the template have as many as 3 powers of a particular momentum in the denominator, in the squeezed limit (e.g., $k_1\equiv k_l\ll k_2\approx k_3\equiv k_s$) the entire template scales like:
\be
\lim_{k_l\ll k_s}B_{\rm equil} \propto \frac{1}{k_lk_s^5}+\dots
\ee

To write an expression for a non-Gaussian field $\Phi$ (the Bardeen potential after reheating) with a bispectrum that matches the equilateral template we use a series of nonlocal functionals of a Gaussian random field $\phi(x)$ \cite{Scoccimarro:2011pz, Baytas:2015nja}:
\be\label{ansatz}
\Phi[\phi(\vv{x})]=\phi(\vv{x})+ f_{\rm NL} \Phi_2[\phi(\vv{x})] +\dots
\ee
where the subscript on $\Phi_2$ indicates the term is quadratic in the Gaussian field. Including higher order terms $\Phi_n$ would add tree-level $(n+1)$ point functions to the model. Since a quadratic term that is local in the Gaussian field, $\Phi_2(x)=[\phi(x)]^2$, gives a bispectrum that diverges as $1/k_l^3$ in the squeezed limit, we must allow $\Phi_2(x)$ to be non-local in order to reduce the degree of divergence. So, we consider
\be
\label{eq:general_ansatz}
\Phi_2(x)= \partial^{\rm \alpha_{\rm 3}}(\partial^{\rm \alpha_2} \phi \partial^{\rm \alpha_1} \phi)(\vv x)\;,
\ee
where the $\alpha_i$ can be negative and the derivative is defined by
\be
\partial^n\phi(x)\equiv\int\frac{d^3k}{(2\pi)^3}k^n\phi(\kv)e^{i\kv\cdot\xv}\;.
\ee
We will restrict consideration to scale-invariant bispectra, which imposes the requirement $\alpha_1+\alpha_2+\alpha_3=0$.

To find the form of $\Phi_2$ that generates a bispectrum for $\Phi$ matching the equilateral template in Eq.(\ref{eq:scaleinv_equil}), notice that there are individual terms in the template that diverge as $1/k_l^3$, $1/k_l^2$, and $1/k_l$. Each of these terms can be recovered from the ansatz Eq.(\ref{eq:general_ansatz}) by considering $|\alpha_i|\leq2$. In addition, individual terms that diverge more strongly than $1/k_l^3$ can be avoided by requiring $\alpha_1, \alpha_2>0$. The quadratic term that satisfies those restrictions is \cite{Scoccimarro:2011pz, Baytas:2015nja}
\bea
\label{eq:Phi_quad}
\Phi_2[\phi(x)] = [ a_1\phi^2+a_2\partial^{-1}(\phi\partial\phi)+a_3\partial^{-2}(\phi\partial^2\phi) \nn \\
+a_4\partial^{-2}(\partial\phi)^2] - {\rm [E.V.]} \;.
\eea
where $-[{\rm E.V.}]$ indicates that the expectation values of the terms should be subtracted. 

To see the bispectrum generated by this quadratic term, it is easiest to first write the general non-Gaussian field in Fourier space (Fourier transform of Eq.(\ref{ansatz})): 
\bea
\label{kspacePhi}
\Phi(\vv k) = \phi(\vv k)+\frac{f_{\rm NL}}{2!}\int\dthreee p1\int d^3p_2 \,[\phi(\vv p_1)\phi(\vv p_2) \nn \\
-\langle\phi(\vv p_1)\phi(\vv p_2)\rangle]\,N_2(\vv p_1, \vv p_2,\vv k)\delta^{3}(\vv k-\vv p_1-\vv p_2)
\eea
where the kernel $N_2$ is symmetric in the first two entries. Then the bispectrum is easily computed:
\be
B(k_1,k_2,k_3)=f_{\rm NL}P_{\phi}(k_1)P_{\phi}(k_2)N_2(k_1,k_2,k_3)+\mbox{perm.}\;
\label{bispectrum}
\ee
where $P_{\phi}(k)$ is the power in the Gaussian field. For the quadratic term in Eq.(\ref{eq:Phi_quad}), the kernel is
\be
N_2(p_1,p_2,k) = 2a_1+a_2\frac{p_1+p_2}{k}+a_3\frac{p_1^2+p_2^2}{k^2}+2a_4\frac{p_1 p_2}{k^2}\;
\label{N_2}\;.
\ee
Not every part of the parameter space covered by the four-parameter kernel $N_2$ in Eq.(\ref{N_2}) above is well behaved. In particular, consider the power spectrum including the (1-loop) non-Gaussian contribution:
\bea
\langle\Phi(k)\Phi(k^{\prime})\rangle = (2\pi)^3\delta(\kv+\kv^{\prime}) \bigg[P_{\phi}(k)+2 \fnl^2 \nn \\
\times \int \! d^3p \, P_{\phi}(p) \, P_{\phi}(|\kv - \vv p|) \,  N_2^2(p,|\kv - \vv p|, k)\bigg]
\eea
The last two terms in the kernel, Eq.(\ref{N_2}), scale like $(p/k)^2$ in the limit $p\gg k$ and so give loop contributions that diverge in the UV. This divergence can be removed by requiring $a_3+a_4=0$. After insisting on good behavior of the loop term (logarithmic divergences only), bispectra from the kernel $N_2$ can still have squeezed limits that diverge like $k_l^{-3}$, $k_l^{-2}$, or $k_l^{-1}$. Restricting to bispectra that are consistent with the single-clock degree of divergence requires that the coefficients of both the $k_l^{-3}$ and the $k_l^{-2}$ contributions vanish. These conditions lead a unique solution, which (with a choice of normalization) corresponds to the equilateral template
\bea \label{equil}
{\rm Equilateral\; Bispectrum:}\;\;\;\;\;\; \nn \\ 
a_{1} = -3, \, a_{2} = 4, \, a_{3} = 2, a_{4} = -2\;.
\eea
Notice that terms $a_2$ and $a_3$ give identical contributions to the shape of the bispectrum. However, they differ in their contributions to the 1-loop power spectrum and so both terms are required to both recover the equilateral template and keep the power spectrum well-behaved.

\subsection{Obtaining the single-clock orthogonal template}
The equilateral template is not sufficient to cover the space of single-clock inflation models that are equally natural within even the simplest effective field theory of single-clock fluctuations \cite{Cheung:2007st}. A second template with squeezed limit proportional to $1/k_l$ but distinct from the equilateral template in other momentum configurations was proposed in \cite{Senatore:2009gt}. (See also the proposal in \cite{Creminelli:2010qf}.) We will show the orthogonal template in Eq.(\ref{eq:orthotemp}) below, but the most important point for our purposes in this section is that the orthogonal bispectrum contains terms with up to four powers of any single momentum in the denominator, i.e., $k_1/k_2^3k_3^4$. This suggests that quadratic terms in the field expansion with a less restrictive set of $\alpha_i$ may generate bispectra with the single-clock squeezed limit but otherwise very little overlap with the equilateral template. Allowing $|\alpha_3|\leq 3$ does not generate any new solutions, but $|\alpha_3|\leq 4$ will be sufficient. 

For our goal of searching for new bispectra consistent with single-clock, we can of course consider $|\alpha_3|$ arbitrarily large. However, as we will see below, even at the same order necessary to recover the orthogonal template we will already find one new shape. For this paper, we will restrict our analysis to considering in detail this new shape and its relationship to models in the literature.

To derive additional factorizable bispectra with the single-clock squeezed limit, we add the following quadratic terms to those in Eq.(\ref{eq:Phi_quad}):
\begin{align}
\label{minimal_set}
&a_5 \partial^{-3}(\phi \partial^{3} \phi), \,\, a_6 \partial^{-3}(\partial \phi \partial^{2} \phi)\nonumber\\
 &a_7 \partial^{-4}(\phi \partial^{4} \phi),\,\, a_8 \partial^{-4}(\partial^{2} \phi \partial^2 \phi), \,\, a_9 \partial^{-4}(\partial \phi \partial^{3} \phi)\nonumber\\
 &a_{10} \partial^2 (\partial^{-1} \phi)^2, \,\, a_{11} \partial^{-4}(\partial^{-1} \phi \partial^5 \phi)
\end{align}
The terms in the last line come from allowing $\alpha_1, \alpha_2\geq-1$. When we restricted to $|\alpha_3|\leq 2$, allowing $\alpha_1, \alpha_2<0$ only generated terms in the bispectrum that were the same as those generated by the four terms in Eq.(\ref{eq:Phi_quad}) (or that were more divergent than $1/k^3$). And, since negative values of $\alpha_1, \alpha_2$ lead to terms with 1-loop contributions to the power spectrum that diverge in the infra-red rather than the UV those terms also would not change the loop constraints. So, including them would have been redundant. However, because the power spectrum goes like $1/k^3$, when we are interested in terms in the bispectra that go as four or more powers of a single momenta the $\alpha_1, \alpha_2<0$ terms do give new functions of momenta at this order. (Table (\ref{table:Kn_operators}) in the Appendix shows a complete list of quadratic terms, including $\alpha_1, \alpha_2<0$ organized by the bispectra they generate to illustrate these redundancies.)

Now, in this new set we need to introduce new constraints which avoid the divergences in the one-loop correction to the power spectrum

\begin{align}
\label{constr}
a_5 + a_6 &= 0, \nonumber\\
a_7 + a_8 + a_9 + a_{11} &= 0\nonumber\\
2a_{10}+a_{11} &=0\;.
\end{align}
where the first two conditions ensure the UV divergence in the loop integral is no more than logarithmic and the last condition does the same for the IR divergence.
Finally, after again imposing that in the squeezed limit the bispectrum diverges no more strongly than $1/k_l$, there are four linearly independent (and non-trivial) solutions for the $\{a_i\}$
\begin{align}
 \label{eq1}
\{a_i\}= \{\{-3,4,2,-2,0,0,0,0,0,0,0 \}, \nn \\
\{1,0,0,0,0,0,-2,2,0,0,0 \}, \nn \\
\{1,-1,0,0,0,0,-1,0,1,0,0 \}, \nn \\
\{1,0,0,0,0,0,-2,0,0,-1,2 \}\} \nn \\
\end{align}

The kernels $N^{(i)}_{2}$, each corresponding to the $i^{\rm th}$ solution from the list above, are
\begin{align}
\label{kernels}
N^{(1)}_{2}(k_1,k_2,k_3) &= -6 + 4\frac{k_1+k_2}{k_3} + 2\frac{k_1^2+k_2^2}{k_3^2} -4 \frac{k_1 k_2}{k_3^2}\nonumber\\
N^{(2)}_{2}(k_1,k_2,k_3) &= 2 - 2\frac{k_1^4+k_2^4}{k_3^4} + 4\frac{k_1^2 k_2^2}{k_3^4}\nonumber \\
N^{(3)}_{2}(k_1,k_2,k_3) &= 2 - \frac{k_1 + k_2}{k_3} - \frac{k_1^4 + k_2^4}{k_3^4} + \frac{k_1 k_2^3 + k_2 k_1^3}{k_3^4}  \nonumber \\
N^{(4)}_{2}(k_1,k_2,k_3) &= 2 - 2 \frac{k_1^4+k_2^4}{k_3^4} - 2 \frac{k_3^2}{k_1 k_2} + \frac{2}{k_3^4}\left(\frac{k_2^5}{k_1} + \frac{k_1^5}{k_2}\right)
\end{align}
where the first line ($N^{(1)}_{2}$) is again the kernel for the equilateral template. Notice that the fourth kernel is actually less divergent than the equilateral template and is just proportional to $1/k_s^6$ in the squeezed limit. We label the bispectra generated by each of these kernels as
\be
B_{i}(k_1,k_2,k_3)\propto P_{\phi}(k_1)P_{\phi}(k_2)N^{(i)}_2(k_1,k_2,k_3)+\mbox{perm.}
\ee

Now we can show that this space of solutions is sufficient to cover the standard orthogonal template \cite{Senatore:2009gt}:
\begin{widetext}
\begin{align}
\label{eq:orthotemp}
\frac{B_{ortho}(k_1,k_2,k_3)}{\fnl^{orth} A_{\Phi}^2(k)} = (1+p)\Bigg[ 6\bigg(\frac{1}{k_1 k_2^2 k_3^3} + \mathrm{5 \, perm.}\bigg) - 6\bigg(\frac{1}{k_1^3 k_2^3} +  \mathrm{2 \,perm.} \bigg) - \frac{12}{k_1^2 k_2^2 k_3^2} \Bigg] - \frac{6\,p}{27} \Bigg[-\bigg(\frac{k_1^2}{k_2^4 k_3^4} + 2 \, \mathrm{perm.}\bigg) \nn \\
+ 6\bigg(\frac{k_1}{k_2^3 k_3^4} + 5 \, \mathrm{perm.}\bigg) - 15 \bigg(\frac{1}{k_1^2 k_2^4} + 5 \, \mathrm{perm.}\bigg) 
-18 \bigg(\frac{1}{k_1^3 k_2^3} + 2 \, \mathrm{perm.}\bigg) + 20 \bigg(\frac{1}{k_1 k_2 k_3^4} + 2 \, \mathrm{perm.} \bigg) \nn \\
+ 12 \bigg(\frac{1}{k_1 k_2^2 k_3^3} + 5 \, \mathrm{perms}.\bigg) + \frac{6}{k_1^2 k_2^2 k_3^2} \Bigg]
\end{align}
\end{widetext}
where $p \simeq 8.52$. The first term in square brackets (multiplied by the factor $(1+p)$), is just the usual equilateral template and the second term with a factor of $p$ itself has the correct single-clock behavior in the squeezed limit. This template can be expressed in terms of the bispectra generated by the kernels we found above:
\bea
B_{ortho} = \bigg(\frac{10p}{9} + 1\bigg) B_{1} - \bigg(\frac{10 p}{9}\bigg) B_{2} \nn \\
+ \bigg(\frac{10 p}{3}\bigg) B_{3} + \bigg(\frac{p}{9}\bigg) B_{4}
\eea

\begin{figure}[htb]
\centering
\begin{minipage}{.45\linewidth}
  \includegraphics[width=\linewidth]{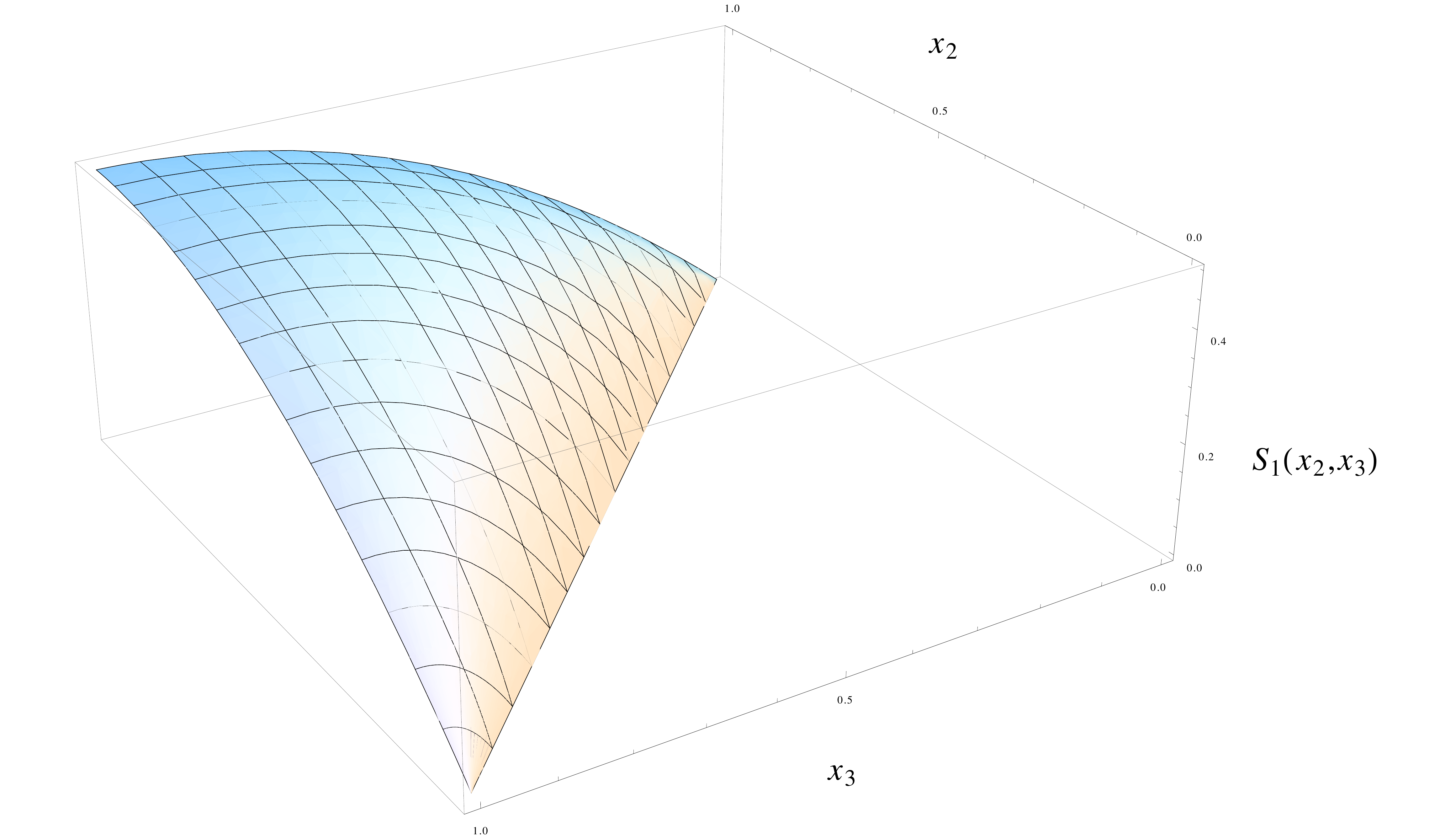}
  \captionof{figure}{$S_{1}(x_2,x_3) \propto (k_1 k_2 k_3)^2 B_{1}(k_1, k_2, k_3)$}
\end{minipage}
\hspace{.05\linewidth}
\begin{minipage}{.45\linewidth}
  \includegraphics[width=\linewidth]{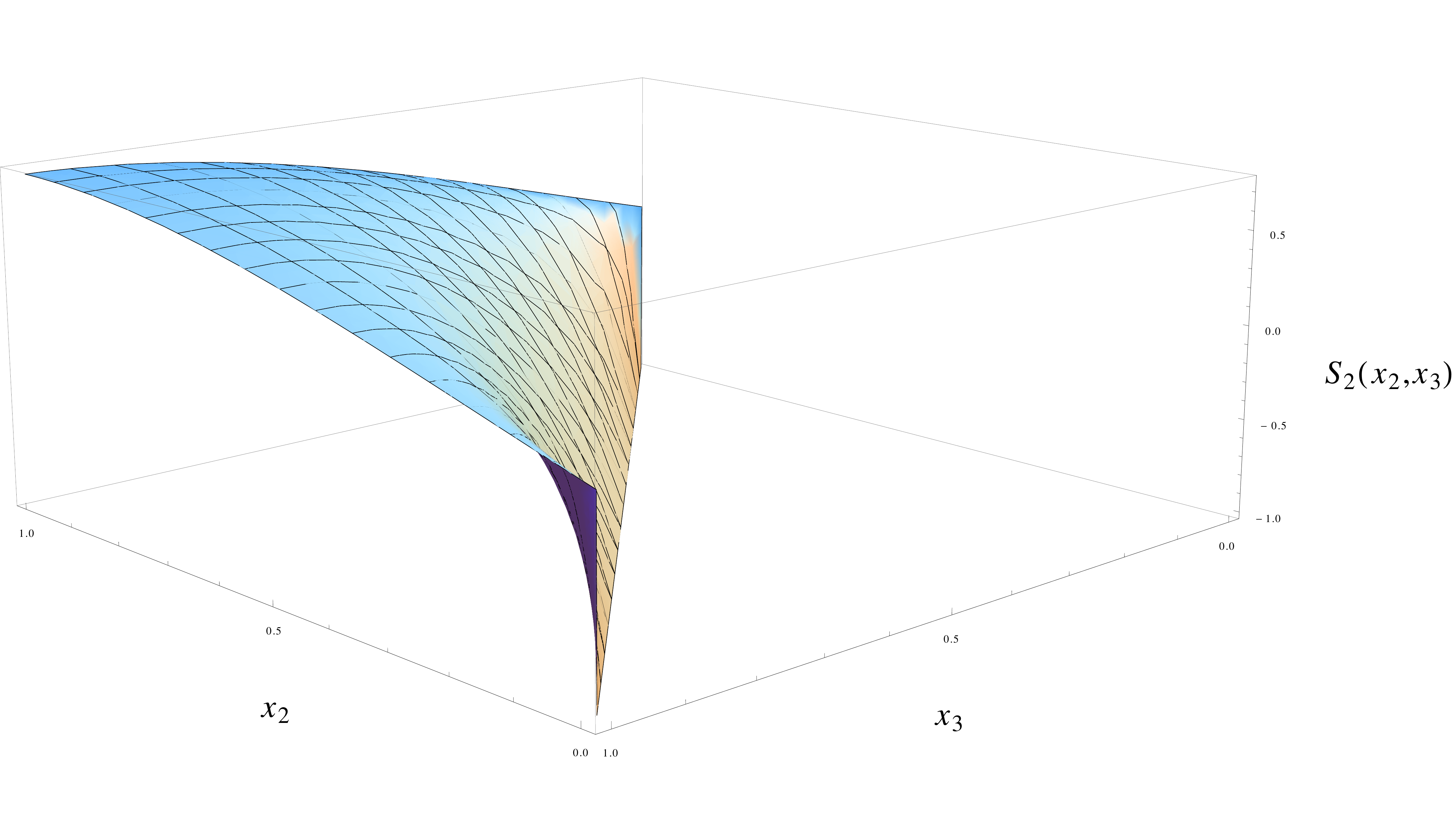}
  \captionof{figure}{$S_{2}(x_2,x_3) \propto (k_1 k_2 k_3)^2 B_{2}(k_1, k_2, k_3)$}
\end{minipage}
\begin{minipage}{.45\linewidth}
  \includegraphics[width=\linewidth]{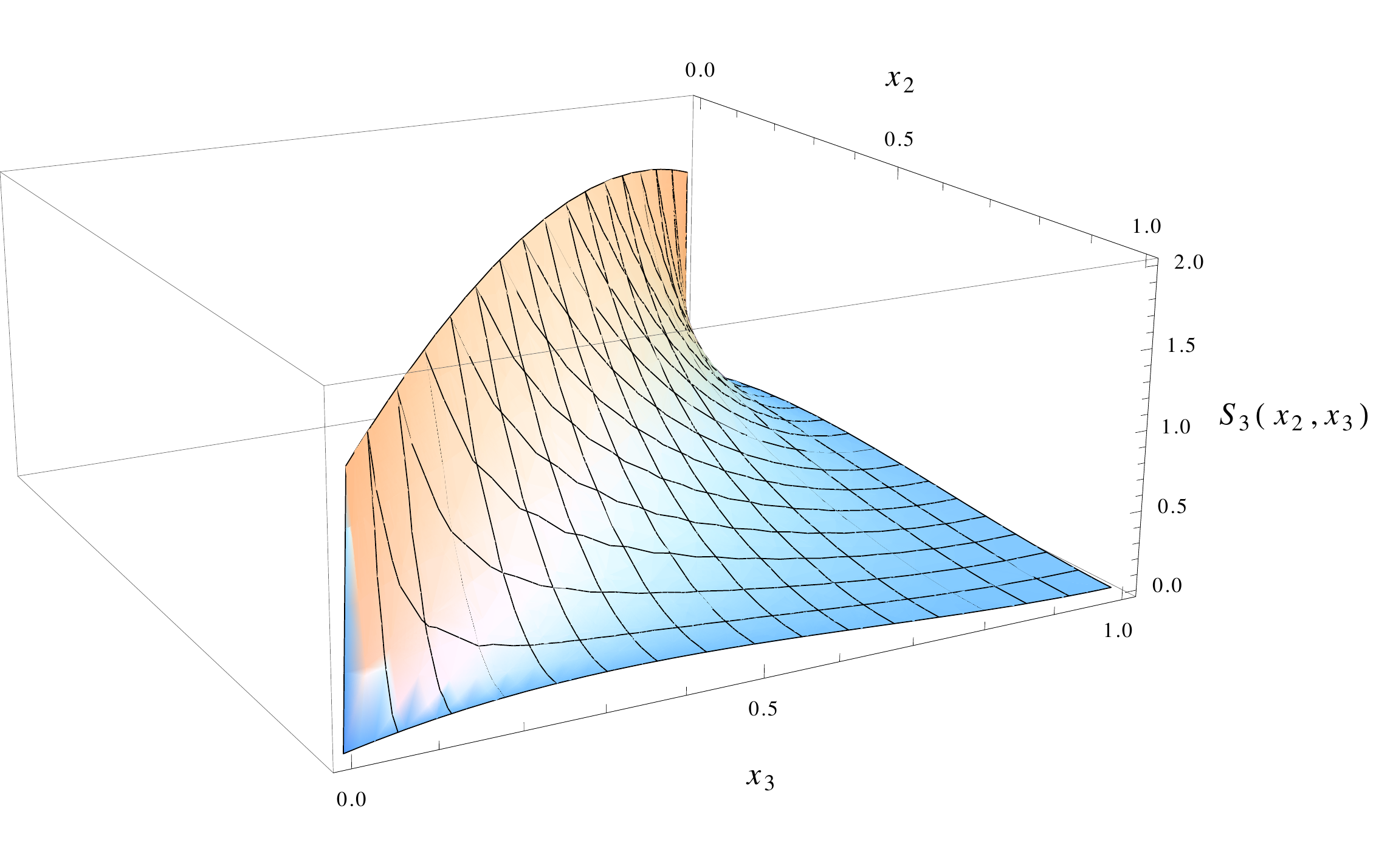}
  \captionof{figure}{$S_{3}(x_2,x_3) \propto (k_1 k_2 k_3)^2 B_{3}(k_1, k_2, k_3)$}
\end{minipage}
\hspace{.05\linewidth}
\begin{minipage}{.45\linewidth}
  \includegraphics[width=\linewidth]{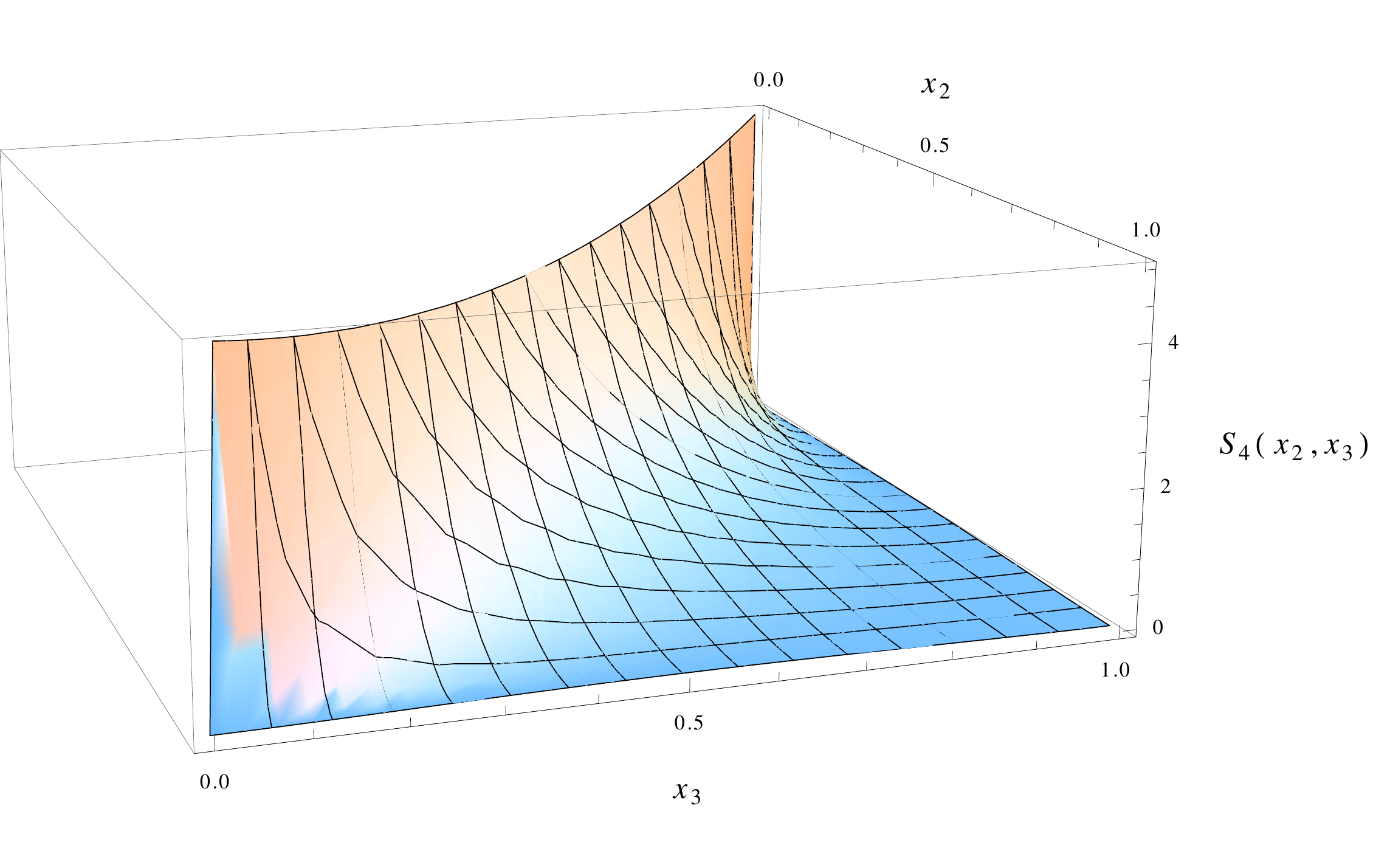}
  \captionof{figure}{$S_{4}(x_2,x_3) \propto (k_1 k_2 k_3)^2 B_{4}(k_1, k_2, k_3)$}
\end{minipage}
\end{figure}

At this point, we have sufficient terms to recover the two standard templates, but do we have anything new? To answer this question we need to define the overlap between two bispectra. This depends on the similarity between the dimensionless ``shapes", $S(k_1,k_2,k_3)$, defined by
\bea
\label{shape}
S_i(k_1,k_2,k_3) =  (k_1k_2k_3)^2 \,\frac{B_i(k_1,k_2,k_3)}{N \fnl A^2_{\phi}} ,
\eea
where $N$ is a normalization factor. Although it is standard in the literature to define $N$ in the equilateral limit, here we are insisting only on a non-vanishing {\it squeezed} limit, and in fact two of our kernels as defined above vanish in the equilateral limit. So, we instead normalize in the squeezed limit. The choice of normalization is, however, irrelevant for comparing any two terms and is irrelevant for expressions involving only the bispectrum. The exact expression for each shape is given in the Appendix, Eq.(\ref{templates_new}), and they are plotted in Figures 1-4.

We can now compare two shapes using the cosine introduced by \cite{Babich:2004gb}: 
\bea
\cos \theta = \frac{S_1 \cdot S_2}{(S_1 \cdot S_1)^{1/2}(S_2 \cdot S_2)^{1/2}},
\eea
where the dot product can be simplified to a two dimensional integral (from three) as the shape is invariant under rescaling of $K=k_1+k_2+k_3$. Defining $k_2/k_1 = x_2$ and $k_3/k_1 = x_3$ gives
\be
S_1 \cdot S_2 = \int_{\texttt{triangle}} dx_2 dx_3 \,\, S_1(x_2,x_3)\,S_2(x_2,x_3)\;.
\ee
The domain of integration can be restricted using rotational invariance and the triangle inequality. Since divergences occur in some shapes at the boundary of the parameter space (when $x_2=0$ or 1) we consider a slightly restricted domain,  $(0.01 \leq x_2 \leq 0.99, \,\, 1-x_2 \leq x_3\leq 1)$. Table \ref{cos_table} shows the correlation coefficients between our four shapes. (Note that if one is interested in a more specific question like the degree to which the bispectrum measured from the cosmic microwave background can distinguish shapes, the definition of the scalar product should be modified with an appropriate weighting function in the integral \cite{Fergusson:2008ra}.)

\begin{table}[htbp]
 \label{table1}
\begin{center}
\begin{tabular}{|c|c|c|c|c|c|}
\hline 
& $S_1$ & $S_2$ & $S_3$ & $S_4$ \\ 
\hline 
$S_1$ & $1$ & $0.74 $ & $0.38$ & $0.30$ \\ 
\hline 
$S_2$ &  & $1$ & $-0.33$ & $-0.41$ \\ 
\hline 
$S_3$ &  &  & $1$ & $0.99$ \\
\hline 
$S_4$ &  &  &  & $1$ \\
\hline  
\end{tabular} 
\caption{The overlap between the shapes associated with the four linearly independent kernels with the single-clock squeezed limit.
\label{cos_table}
}
\end{center}
\end{table}

\begin{figure}[htbp]
\centering
\begin{minipage}{.55\linewidth}
  \includegraphics[width=\linewidth]{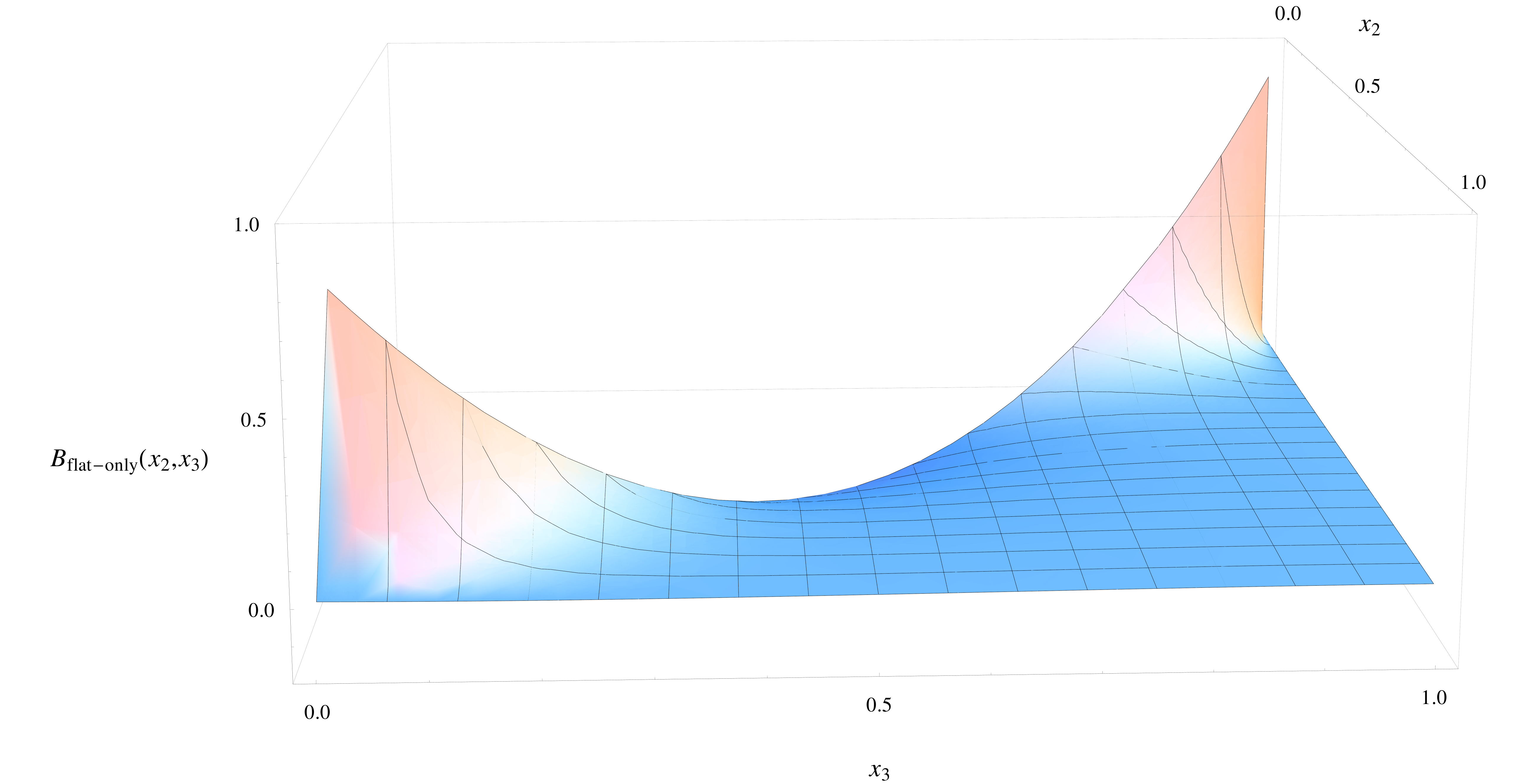}
  \captionof{figure}{This is the 3D plot of the flat-only shape $S_{flat-only}(x_2,x_3) \propto (k_1 k_2 k_3)^2 B_{flat-only}(k_1, k_2, k_3)$\label{fig:new3D}}
\end{minipage}
\hspace{.05\linewidth}
\begin{minipage}{.35\linewidth}
  \includegraphics[width=\linewidth]{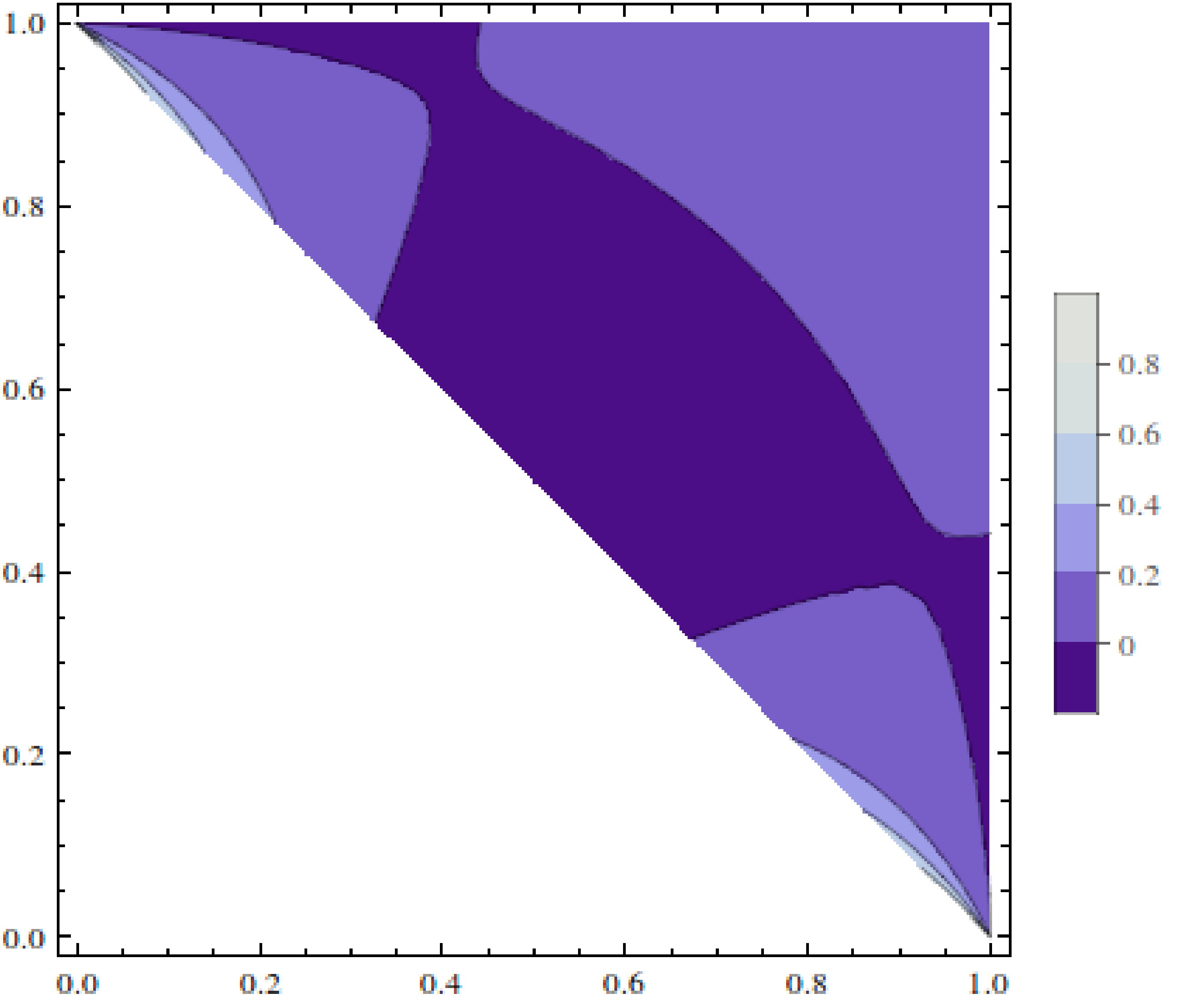}
  \captionof{figure}{Contour plot of the flat-only shape.\label{fig:newContour}}
\end{minipage}
\end{figure}

The table shows significant overlap between $S_3$ and $S_4$, so we expect that our basis includes just one additional shape with very little overlap with the standard equilateral and orthogonal templates. We express this new shape by searching for a linear combination of the shapes associated to $B_1$, $B_2$, and $B_3$ with minimal cosine with both the equilateral and orthogonal templates. This procedure gives $ \cos(B_{new},B_{ortho}), \, \cos(B_{new},B_{equil}) \} \lesssim \mathcal{O}(10^{-2})$ for 
\begin{align}\label{new}
B_{flat-only} = 0.486 \, B_1 - 0.484 \, B_2 + 0.998 \, B_3
\end{align} 
(We have written this in terms of the bispectrum rather than the shape functions, so it is independent of normalization convention for the individual shapes associated with the $B_i$. To understand the shape above qualitatively, we plot it in both 3D (Figure \ref{fig:new3D}) and as a contour plot (Figure \ref{fig:newContour}).

The graph peaks along flattened configurations $x_2+x_3=1$ and has essentially very small amplitude away from that line, including at the equilateral point. We label it ``flat-only" to emphasize both where it peaks and the small equilateral amplitude. Below, we will compare this shape in detail with a variety of single-clock models, but the fact we will find low overlap is perhaps already not surprising: because the shape is nearly zero away from the flat limit, its cosine with other shapes is predominantly a function of the behavior only along the flat limit. For reference in the next section, note that the oscillation along the line $x_2+x_3=1$ can be well approximated by
\bea
S_{\rm flat-only}(1,x_2,x_3=1-x_2)\approx 0.86 \nn \\
- {\rm Sin}\left[(1+x_2-x_3)\left(\frac{\pi}{2}\right)\right]\;.
\eea
This simple oscillatory behavior turns out not to be an easy feature to find. For reference, the cosine of the flat-only shape with the local shape is about 0.5.

\subsection{Comparison with previous work}
Although our approach and motivation were rather different, it is interesting to note that our results can be mapped to terms in the basis constructed by Byun and Bean \cite{Byun:2013jba}. We have constructed a non-Gaussian field while they focused on a basis for bispectra. The system of constraints we are solving is different since we have several field terms that produce the same polynomials in the bispectrum, and additional constraints on the behavior of UV loops. In both their basis and ours, one has to choose some order to truncate the list of terms considered, and truncating at the appropriate order to recover the single-clock orthogonal template turns out to give a set of four linearly independent solutions in either case. Appendix \ref{basis} shows in detail the relationship between these approaches and the results. Byun and Bean similarly computed cosines and identified the existence of a shape ($S_4^{\rm BB}$ in Appendix \ref{basis}) that had cosines of 0.07 and 0.8 with the equilateral and orthogonal templates, respectively. So, this shape is not well covered by the two standard templates. Although $S_4^{\rm BB}$ itself only has cosine of about 0.1 with the flat-only shape, it is an essential ingredient in the linear combination of the Byun and Bean basis that is equivalent to the flat-only shape.  

In the next section, we examine in detail the relationship of this flat-only shape to the shapes that can be produced by single-clock Lagrangians, including some natural shapes proposed after \cite{Byun:2013jba} appeared.

\section{Comparing to single-clock inflation}
\label{sec_singleclock}

The bispectra for fluctuations from single-clock inflation can be organized as an effective field theory (EFT) for the fluctuations of the `clock' field about the assumed Friedmann-Robertson-Walker background metric. The simplest version of this effective theory was originally proposed in \cite{Cheung:2007sv} and can be extended to impose additional symmetries on the fluctuations (eg, Galileon symmetry) \cite{Burrage:2010cu, Creminelli:2010qf} or to consider cases where slightly more complex physics may be relevant. Then one may consider, for example, higher order extrinsic curvature terms \cite{Bartolo:2010bj}, a discrete rather than continuous symmetry for the fluctuations (resonant non-Gaussianity) \cite{Hannestad:2009yx, Flauger:2010ja, Chen:2010bka, Behbahani:2011it}, dissipative effects in the inflationary background \cite{LopezNacir:2011kk} or the presence of more than one scale suppressing the higher derivative terms \cite{Behbahani:2014upa}. (Extensions that break the single-clock nature of the model, for example allowing non-Bunch-Davies initial states \cite{Agarwal:2012mq} or to include some possible multiple field scenarios \cite{Senatore:2010wk} are also possible.) The various single-clock scenarios generate bispectra which all share the same squeezed limit scaling but can be sufficiently different functions away from that point in momentum space that they can be observationally distinguished. We begin by reviewing the minimal EFT, which is well covered by the equilateral and orthogonal templates alone. We then consider the additional bispectra that have been proposed by allowing richer, but still effectively single clock, physics to see where the folded-only shape could be generated. For more detailed discussions and derivations of single-clock actions and the symmetry arguments underlying its structure (emphasizing a variety of points), see also \cite{Creminelli:2012ed, Hinterbichler:2013dpa, Anderson:2014mga, Chluba:2015bqa}.

\subsection{The lowest order terms in the effective theory of perturbations}
The effective field theory of single clock fluctuations assumes a background solution for the metric that is homogeneous and isotropic, but time-dependent, so that it is primarily characterized by the time evolution of the Hubble scale, $\dot{H}$. Since this background solution to Einstein's equations breaks time-translation invariance by assumption, there is a propagating scalar degree of freedom that, in a gauge where the matter sourcing the time evolution is homogeneous on equal-time surfaces, is just a metric fluctuation \cite{Cheung:2007st}. In ``single-clock" models, this adiabatic mode is the only physical scalar fluctuation and can be considered as the Goldstone mode related to the breaking of time translation symmetry. Written as a standard scalar mode of mass dimension one, the fluctuation $\pi$ is related at lowest order to the usual scalar curvature perturbation conserved outside the horizon, $\zeta$, by $\pi = - \zeta/H$. (And $\zeta=5/3\Phi$, where $\Phi$ is the matter era Bardeen potential.) In the limit where all background quantities are slowly evolving, and all higher order derivative terms are suppressed according to dimension, and the coupling to other metric degrees of freedom is negligible, the action for the Goldstone mode up to leading cubic terms is \cite{Cheung:2007st,Senatore:2009gt}:
\bea
\label{eff_action}
S = \int dx^4 \sqrt{-g} \Bigg[-\frac{M_{Pl}^2 \dot{H}}{c_s} \Big(\dot{\pi}^2 - c_s^2 \frac{1}{a^2} (\partial_i \pi)^2 \Big) \nn\\
+ \frac{\dot{H} M_{Pl}^2}{c_s^2}(1-c_s^2)\dot{\pi}\frac{1}{a^2}(\partial_i \pi)^2 \nn \\
- \frac{\dot{H}M_{Pl}^2}{c_s^2}(1-c_s^2)\Big(1+\frac{2}{3} \frac{\tilde{c}_3}{c_s^2}\Big) \dot{\pi}^3 + \cdots \Bigg],
\eea
where $c_s$ is the speed of sound of the fluctuations and $\tilde{c}_s$ is the dimensionless parameter of order of unity. The form of the cubic terms can be understood from Lorentz invariance together with the requirement that the dominant interactions of $\pi$ when the background is nearly time-translation invariant should be those that are invariant under $\pi\rightarrow\pi+c$ (so every $\pi$ field should carry a derivative). Furthermore, each $\pi$ carries only one derivative; terms with more derivatives should be suppressed by powers of $H/\Lambda$ where $\Lambda$ is a mass scale $>H$. 

Since there are two independent terms in the Lagrangian above, it is natural to expect that the space of linear combinations of those two terms can be described by (at most) two independent templates. It will be important for the rest of our discussion to note that the bispectra from the individual operators are not necessarily the orthogonal elements of the basis. In fact, in the Lagrangian above the three-point functions generated by the individual cubic operators $\dot{\pi}(\partial_i \pi)^2$ and $\dot{\pi}^3$ both have high overlap with the equilateral template (cosines $\gtrsim0.9$) and with each other (cosine $\sim0.97$), but low overlap with the orthogonal template (cosines $\lesssim0.3$). So, it is possible to find a range of values of $\tilde{c}_3$ where the bispectra from linear combinations of the operators have very low overlap with the equilateral shape. Specifically when $\tilde{c}_3 = -5.4$ (and $c_s\ll1$) the bispectrum is almost exactly the orthogonal shape. In other words, the orthogonal template has been designed to have very small overlap with the equilateral shape, and to cover the `gap' in the single-clock models that have low overlap with the equilateral template \cite{Senatore:2009gt}. Since our new template has very small cosine with both of the orthogonal basis elements that cover the space of shapes corresponding to these operators, we expect it to have low overlap with any linear combination of the operators themselves, and with any linear combination of the two templates. This can be checked explicitly, and Appendix \ref{app:LC} contains further details about cosines between a test shape and linear combinations of reference shapes. 

Clearly, the space of orthogonal shapes that appear equally natural at the lowest order in this effective field theory do not appear at the same order in our basis for the field, or in a basis for bispectra organized around the maximal degree of divergence of individual polynomial terms: the equilateral shape is the unique solution for terms that have no more than three powers of an individual momenta in the denominator, while both the orthogonal and flat-only shapes appear when we allow up to four powers. So, as a first step in searching for the flat-only shape, we might ask if it is generated by higher order operators (or a fine-tuned linear combination of them). This is discussed next, in the context of physical reasons one might be interested in terms with more derivatives.  

\subsection{Beyond the minimal EFT: terms with more derivatives}
\label{comparison}
An important part of the restriction to just two cubic terms above came from the assumption that all higher derivative cubic terms (terms with more than one derivative acting on a $\pi$ field) were suppressed compared to those appearing in Eq.(\ref{eff_action}). However, there may be reasons why some terms with more derivatives may be naturally enhanced. For example, \cite{Creminelli:2010qf}, examined the action that comes from imposing approximate Galilean symmetry instead of a shift symmetry
\bea
\pi \rightarrow \pi + b_{\mu} x^{\mu} + c
\eea
with $b_{\mu}$ and c are constant. The set of terms which are independent and respect the Galilean symmetry are 
\bea
\label{eq:GalOps}
\mathrm{\textbf{Galilean:}} \,\,\,\,\,\, \mathcal{O}_1 = \ddot{\pi}^3, \,\, \mathcal{O}_2 =\ddot{\pi} (\partial_i \dot{\pi} - H \partial_i \pi)^2, \nn \\
\mathcal{O}_3 = \ddot{\pi} (\partial_i \partial_j \pi)^2 - 2 H \dot{\pi} \ddot{\pi}^2 + 3 H^3 \dot{\pi}^3
\eea
There are three independent terms here, which may all have coefficients of the same order. However, we still do not find significant overlap with the flat only shape: the first two operators give bispectral shapes very close to those from the original two cubic operators in Eq.(\ref{eff_action}), and although the third does have a peak in the flat limit at $x_2=x_3=\frac{1}{2}$, it is still very nearly a linearly combination of the equilateral and orthogonal templates \cite{Creminelli:2010qf}. Just for completeness, we check the relationship between the bispectrum generated by these operators, and linear combinations of pairs of the operators, and our flat-only shape and find no significant overlap, $|\cos (B_{flat-only}, B_{\mathcal{O}_{\rm Galilean}})| \lesssim 0.1$.

A second way to enhance the importance of some additional higher derivative cubic terms in the effective action was recently proposed by \cite{Behbahani:2014upa}, and {\it does} require additional template shapes distinct from the orthogonal and equilateral templates. The physics introduced here is to allow two different mass scales ($\Lambda$, $\Lambda'>H$) suppressing higher derivative terms so that the relative size of the terms is not purely a function of number of derivatives. In particular, suppose the general form of the cubic interactions is
\bea
\mathcal{L'}_3=\frac{1}{(\Lambda')^{n-m-1}}\partial^{n-m} \pi^3, \,\, \mathcal{L}_3=\frac{1}{(\Lambda)^{n-1}}\partial^{n} \pi^3
\eea
where $\Lambda'^{n-m-1} > \Lambda^{n-1}/H^m$. Now some terms with more derivatives may generate correlations with an amplitude as large as terms with fewer derivatives. Imposing technical naturalness and searching for the set of operators that generate bispectra with small overlap with the equilateral and orthogonal templates, \cite{Behbahani:2014upa} find that the space of technically natural models can be well covered with the addition of templates corresponding to two new operators:
\bea \label{oper}
\mathrm{\textbf{Two scales:}} \,\,\,\,\,\, \mathcal{O}_1 = \dddot \pi (\partial_i \partial_j \pi)^2, \,\, \mathcal{O}_2 =  \dddot \pi (\partial_i \ddot \pi)^2
\eea
Because of the spatial derivatives in these cubic terms, the resulting shapes are proportional to one or two powers of the dot product of two momenta (eg, $\vec{k}_2 \cdot \vec{ k}_3$) and do have oscillations along the line $x_2+x_3=1$. Using the exact bispectra from these operators, we compute the cosines between these higher derivative shapes and our flat-only template. We find some overlap, but not enough to conclude that we can replace the flat-only template with these shapes:
\bea
cos(B_{flat-only},B_{\mathcal{O}_1}) = 0.36 \nn \\
cos(B_{flat-only},B_{\mathcal{O}_2}) = 0.47
\eea
Essentially, these shapes have too much structure (more turning points) along the flattened limit to be our shape. (See Figure 2 of \cite{Behbahani:2014upa} for plots of these shapes.) Although \cite{Behbahani:2014upa} did not give an explicit form for factorizable templates for the shapes above, this result also means that we expect to recover these shapes only if we extend our formalism to allow even higher powers of inverse derivatives. 

The two shapes above were chosen to be relatively orthogonal to the each other and to the equilateral and orthogonal shapes, so we do not expect that considering linear combinations will lead to a higher cosine. Indeed, we have checked that the flat-only shape has small overlap with all linear combinations of pairs of operators from the set of the equilateral, orthogonal, and the two higher derivative shapes here. Although we have not checked if there is any physical motivation for considering this case, we note that the highest cosine we find ($\approx 0.57$) is with a linear combination of $\mathcal{O}^{\rm Gal.}_3$ from Eq.(\ref{eq:GalOps}) and $\mathcal{O}^{\rm high.\,deriv.}_2$ from Eq.(\ref{oper}).

We can also check if the flat-only shape arises from higher derivative terms without worrying about whether the amplitude is naturally enhanced or not. Indeed, there are some higher derivative operators that are known to produce bispectra that peak in the folded configuration. For example, as emphasized by \cite{Bartolo:2010bj} the set of operators 
\begin{align}
\dot{\pi} (\partial_{i} \partial_j \pi)^2, \, \partial^2_i \pi (\partial_j \partial_k \pi)^2, \,
(\partial_{i} \partial_j \pi) (\partial_{j} \partial_k \pi) (\partial_{k} \partial_i \pi) \nn
\end{align}
which are related to variations in the extrinsic curvature of equal time surfaces, generate bispectra have high overlap with the ``flattened" template \cite{Creminelli:2010qf}
\begin{align}
B_{flattened} \propto 0.6 \bigg[\frac{16}{9 k_1 k_2 k_3^4} + \frac{k_1^2}{9 k_2^4 k_3^4} - \frac{1}{k_1^2 k_2^4} + \mathrm{perms.} \bigg] \nn \\
+ \bigg[\frac{1}{k_1^3 k_2^3} - \frac{1}{k_1 k_2^2 k_3^3} + \mathrm{perms.} \bigg] + 2(1.6)\frac{1}{k_1^2 k_2^2 k_3^2} \;.
\end{align}
Actually, even a special choice of parameters in the simplest EFT ($\tilde{c}_3=-6$) is also well-approximated by this template. That means this template can be expressed as a linear combination of the equilateral and orthogonal templates and so it is not surprising to find small overlap with the flat-only shape:
\bea
cos(B_{flat-only},B_{flattened}) = 0.003\;\nn
\eea
Again, this shape is in the family covered by the local and orthogonal templates, so considering it with linear combinations of the other shapes above should not lead to high overlap with the flat-only template.

\subsection{Beyond the minimal EFT: terms suppressed by time-dependent coefficients}
In the previous sections we considered terms in the action for fluctuations that were precisely invariant under a shift $\pi\rightarrow\pi+c$ by including only terms with at least one derivative on each $\pi$ field and by assuming the coefficients of each term were exactly constant. In general, this symmetry is weakly broken and it is natural to expect some time-dependence in the coefficients, and some terms with no derivatives acting on $\pi$. However, the amplitude of these terms should in general be suppressed by slow-roll parameters. Their signal can be maximized by imposing physics where the action for $\pi$ must instead remain invariant under a discrete shift symmetry, $\pi(t,\vec{x})\rightarrow\pi(t,\vec{x})+2\pi/\omega$. Now although it is still true that terms with no derivatives acting on $\pi$ must be multiplied by time derivatives of the background evolution, those terms can scale with factors of a large parameter $\omega/H\gg1$. The discrete symmetry gives rise to ``resonant" effects in the correlation functions \cite{Hannestad:2009yx, Flauger:2010ja, Chen:2010bka, Behbahani:2011it, Chen:2008wn} as the amplitude of Fourier modes is enhanced as they are stretched through $k/a(t)\approx\omega$. As the name implies, the bispectra from models in this class have oscillatory features and are scale-dependent, but still satisfy the consistency condition for sufficiently squeezed configurations. Our flat-only shape is scale-invariant and only has oscillations along the flat line (with $K=k_1+k_2+k_3$ constant). Still, could it be a good approximation to a more complex shape  \cite{Gwyn:2012pb}?

To see that the answer is no, note that the dominant terms in the resonant shape are 
\bea
 S_{res}(k_1,k_2,k_3) = \sin\bigg(\alpha\,{\rm ln}(K/k_*)\bigg) \nn \\
 + \frac{1}{\alpha}\cos\bigg(\alpha\,{\rm ln}(K/k_*)\bigg) \sum \limits_{i \neq j} \frac{k_i}{k_j}\;,
\eea
where $\alpha(=\omega/H)$ sets the wavelength of the oscillation in $\log(k)$ space and the reference scale $k_*$ effectively sets the phase. Note that the resonant shape depends only on the scalar sum of momenta $K \equiv k_1+k_2+k_3$ and so does not contain any oscillation along the folded line $x_2+x_3=1$, $K=$constant. We find little overlap between our folded-only shape and the resonant shape for a range of parameter values.

The physics that gives this dominant resonant shape also causes deviations from the Bunch-Davies vacuum that generically contains terms that depend on $\tilde{k}_j=K-2k_j$ (as do more general bispectra from models with small sound speed together with deviations from non Bunch-Davies initial states \cite{Meerburg:2009ys,Meerburg:2009fi}). In principle, these pieces could have larger overlap with the folded-only shape, although they are again scale-dependent and need not satisfy the single-field consistency relation on intermediate scales. The proposed template for folded resonant non-Gaussianity is \cite{Chen:2010bka}
\begin{widetext}
\begin{align}
S_{\rm fold.\,reson.}=e^{-C^{3/5}(1+x_2-x_3)/2} \sin\left[C\left(\frac{(1+x_2-x_3)}{2}+\ln(x_3/x_*)\right)+\phi\right] + 2\,{\rm perm.}
\end{align}
\end{widetext}
However, varying parameters in this ansatz we again find no more than modest overlap with the flat-only shape, $C(S_{flat-only},S_{\rm fold.\,reson.})\sim0.3$. We have included this shape with two parameter choices (one chosen to maximize its individual cosine with the flat-only shape, the other to match the central oscillatory feature along the flat line) in checks of linear combinations (together with the three Galilean operators and the seven and nine-derivative operators from the previous section). We find that adding these shapes does not increase the cosines among this set any further. The folded resonant template has more structure in the corners of the parameter space (i.e, $x_2=0,1$), and away from the flat limit, than the flat-only template does, and we suspect this is driving the small cosine. (There is also small overlap with the `enfolded template' \cite{Meerburg:2009ys}, which does not have the single-clock squeezed limit, and does not have an oscillation along the flat line.) 

Finally, we have checked the cosine between the flat-only shape and 35 higher order terms that contain one $\pi$ with no derivatives, finding cosines generally in the range of $0.5-0.6$, but we have not examined all linear combinations to rule out the possibility that flat-only shape is a fine-tuned linear combination of some operators.

\subsection{Beyond the minimal EFT: dissipative effects}
Although the flat-only shape could still be fine-tuned member of the space of operators considered in the previous two subsections, it may also be that its structure is a clue that it can only be naturally generated by some physics that is a more non-trivial change to the original EFT. One example of such physics was considered by \cite{LopezNacir:2011kk}, who added dissipative effects to the effective field theory, generated by extra degrees of freedom that do not change the single-clock nature of the model \cite{LopezNacir:2012rm} but that do change the dynamics of inflation by directly coupling to the clock around or before the time the modes we observe cross the horizon. Although \cite{LopezNacir:2011kk} did not systematically study all possible non-Gaussianities from this rich class of models, the examples they did examine in detail peak on the equilateral configuration and around $x_2 \simeq x_3 \simeq 1/2$, with no oscillation along the flat limit. We again found only modest overlap comparing to a couple of example shapes from \cite{LopezNacir:2011kk}. If these properties are general, which seems physically reasonable, it seems unlikely that the flat-only shape is a typical member of that class of models. 

\section{Discussion}
\label{sec_discuss}
In this work we have constructed a set of non-Gaussian fields built from non-local real space expressions up to second order in Gaussian field, and by expanding the second order term in the number of inverse derivatives. We restricted to fields with tree-level bispectra that couple short and long wavelength modes no more strongly than single-clock inflation does. Long wavelength background modes from these bispectra just act like curvature and so the coupling introduces no additional cosmic variance in the locally observed power spectrum. At the same order that we recover the familiar equilateral and orthogonal templates, we also find we can construct a third `flat-only' shape with negligible overlap with those two. The shape is characterized by primarily coupling modes in flattened triangle configurations (and very little in the equilateral limit). In addition, it has a simple oscillation along the flat line. 

We systematically investigated physics that can enhance terms beyond the minimal, lowest order shapes in the effective field theory of single-clock fluctuations including multiple ways to enhance terms with more derivatives and terms where one $\pi$ carries no derivatives. We also computed cosines between the flat-only shape and typical suppressed terms with more or fewer derivatives. Finally, we looked at some shapes produced by non-trivial modifications of the background dynamics (dissipation). We found that none of these cases had the structure to naturally provide the oscillation along the flat-only limit that the template has. This all suggests that the flat-only shape may not describe any `natural' single-clock model so far in the literature. We have checked a subset of linear combinations of single-clock shapes, but we have not been exhaustive. The shape could potentially be constructed from a fine-tuned combination of these standard effects, in which case it is not clear whether behavior along the flat limit indicative of any particular physics. It is also possible that the flat-only shape occurs naturally in a model that is not single-clock, but whose bispectrum accidentally realizes the symmetries associated with single-clock inflation. It would be interesting to try to reverse-engineer a model that generates this shape. 

The broader context for this work is a better understanding of how best to frame constraints from data on the space of possible non-Gaussian correlations. We have taken the point of view that the degree to which long modes are coupled to short modes is the most important qualitative feature and used that as an organizing principle. At any order, the space of fields whose bispectra have no appreciable cosmic variance from mode coupling must be greater than or equal to the space of shapes (natural or fine-tuned) from single-clock inflation. The details of the bispectrum away from the squeezed limit carry information about the physics of the model beyond its single-clock nature, and possibly even about whether the single-clock squeezed limit could be accidental. 

Rather than finding a basis for the bispectrum, we have used a basis for the non-Gaussian field. This is useful because the field can be numerically realized. However, in the case where we expand only to the first non-Gaussian term and insist that it generates no cosmic variance in lower order correlation functions, correlations at different orders are only related via loop corrections. And, as we have shown by comparison with the earlier work of Byun and Bean \cite{Byun:2013jba}, there is sufficient degeneracy in going between the bispectrum shape and the field that loop corrections can always be satisfied. In other words, the correlation functions are pretty much decoupled and one may as well construct a basis for each order independently. But, when we expand the space of allowed correlations to include significant mode-coupling, the relationship between correlations at different orders is very informative (think, eg, of the non-Gaussian halo bias) and so constructing the field rather than the bispectrum alone is a better approach. As we showed in \cite{,Baytas:2015nja}, mode coupling from higher order correlation functions can be entirely consistent with the presence of a bispectrum that has the single-clock (no cosmic variance) squeezed limit.

Finally, we note that the existing literature suggests that it would be worthwhile to continue this procedure of constructing the quadratic term to higher order in inverse derivatives. One can continue to compare to the construction of Byun and Bean, and the natural single-clock bispectra proposed in \cite{Behbahani:2014upa} should be recovered. it would be interesting to compare the number total number of orthogonal, super cosmic variance free shapes constructed this way to the number so far discussed in the single-clock inflation literature. 

\acknowledgements We are grateful to Nishant Agarwal and Joyce Byun for very useful comments on a draft of this article, which helped us clarify the presentation. This work was supported by NSF Award PHY-1417385. In addition, we thank the Perimeter Institute for hospitality while this work was in progress. This research was supported in part by Perimeter Institute for Theoretical Physics. Research at Perimeter Institute is supported by the Government of Canada through Industry Canada and by the Province of Ontario through the Ministry of Economic Development \& Innovation.

\appendix

\section{Correspondence with basis functions for bispectra of Byun and Bean}
\label{basis}
Byun and Bean \cite{Byun:2013jba} introduced a set of basis functions, $\{\mathcal{K}_n\}$, that describes nearly scale-invariant bispectral shapes (recall shape $S$ is proportional to $(k_1k_2k_3)^2B(k_1,k_2,k_3)$) with various levels of divergence in the squeezed limit. The $\mathcal{K}_n$ are symmetrized, separable polynomials of momenta
\bea \label{K}
\mathcal{K}_n(k_1,k_2,k_3) \equiv \frac{1}{N_n} \bigg(k_1^{p}k_2^{r}k_3^{s} + perms. \bigg),
\eea
where $n$ labels the particular triple of (positive and negative) integers $\{p,r,s\}$ and $N_n$ is the number of distinct permutations. We will restrict to exact scale-invariance of the power spectrum and bispectrum for simplicity, so $p+r+s=0$. Byun and Bean started labeling with the constant shape
\be
\mathcal{K}_0 = 1
\ee
and then considered sets of polynomials grouped according to the minimum power of momenta appearing, $R=-1,-2,-3,$ etc. There are six additional terms up to $R=-2$ divergence: 
\begin{align}
\mathcal{K}_1 = \frac{1}{6}\bigg(\frac{k_3}{k_1} + 5 \,\, perms \bigg), \,\,\mathcal{K}_2 = \frac{1}{3} \bigg(\frac{k_1^2}{k_3k_2} + 2 perms \bigg) \nn \\
\,\, \mathcal{K}_3 = \frac{1}{3} \bigg(\frac{k_1k_2}{k_3^2} + 2 \,\, perms \bigg), \,\,
\mathcal{K}_4 = \frac{1}{6} \bigg(\frac{k_1^2}{k_3^2} + 5 \,\, perms \bigg) \nn \\
\mathcal{K}_5 = \frac{1}{6} \bigg(\frac{k_1^3}{k_2k_3^2} + 5 \,\, perms \bigg), \,
\mathcal{K}_6 = \frac{1}{3} \bigg(\frac{k_1^4}{k_2^2k_3^2} + 2 \,\, perms \bigg) 
\end{align}
\begin{table*}[hbt] 
\begin{center}
\begin{tabular}{| l | c| c | c | c | l |}
\hline
$\mathcal{K}_n$ &$\{p,r,s\}$ & \multicolumn{3}{c}{ $\Phi_2(x)$} & $\{a_i\}$ \\ 
 \hline\hline
$\mathcal{K}_0$ &$(0,0,0)$&  $\partial^{-2}(\partial \phi)^2$ & -- & -- & $\{a_4$\}\\ 
$\mathcal{K}_1$ &$(-1,0,1)$& $\partial^{-1}(\phi \partial \phi)$ & $\partial^{-2}(\phi \partial^{2} \phi)$ & $\partial^{-3}(\partial \phi \partial^2 \phi)$ & $\{a_2,a_3,a_6\}$ \\ 
$\mathcal{K}_2$ &$(-1,-1,2)$& $\phi^2$ & $\partial^{-3}(\phi \partial^{3} \phi)$ & --& $\{a_1,a_5\}$ \\ 
$\mathcal{K}_3$ &$(-2,1,1)$& $\partial^{-4}(\partial^2 \phi)^2$ & $\partial^{-1}(\partial^2 \phi \partial^{-1} \phi)$ & --& $\{a_8,\times\}$ \\ 
$\mathcal{K}_4$ &$(-2,0,2)$& $\partial^{-4}(\partial^3 \phi \partial \phi)$ & $\partial^{-2}(\partial^{-1} \phi \partial^{3} \phi)$ & $\partial^{-1} \phi \partial \phi $ & $\{a_9,\times,\times\}$ \\ 
$\mathcal{K}_5$ &$(-2,-1,3)$& $\partial^{-4}(\phi \partial^{4} \phi)$ &  $\partial^{-3}(\partial^{-1} \phi \partial^4 \phi)$ & $\partial(\phi \partial^{-1} \phi)$ & $\{a_7,\times,\times\}$ \\ 
$\mathcal{K}_6$ &$(-2,-2,4)$& $\partial^2(\partial^{-1} \phi)^2$ &  $\partial^{-4}(\partial^{-1} \phi \partial^5 \phi)$ & --& $\{a_{10},a_{11}\}$ \\ 
\hline
\end{tabular}
\end{center}
\caption{The set of terms in the real space field expansion, $\Phi_2(x)$ that correspond to each polynomial $\mathcal{K}_n$ (see Eq.(\ref{K})) appearing in the shape of the bispectrum. The different $\Phi_2(x)$ terms on a given horizontal line result in redundant terms in the shape of the bispectrum but contribute differently to the lowest order non-Gaussian (loop) contribution to the power spectrum. The conditions of a well-behaved loop correction in general introduce constraints among terms corresponding to different from $\mathcal{K}_n$. However, some $\Phi_2(x)$ terms are truly redundant even once loops are considered (eg, they come with unconstrained coefficients) and so are not needed. The last column lists the labels $a_i$ we used in the main text, where $\times$ indicates the term was redundant for our purposes and so was not included. \label{table:Kn_operators}}
\end{table*}
Byun and Bean then find four linear combinations of these shapes that have the same squeezed limit as the equilateral template
\begin{align}
S^{\rm BB}_1 &= -2\mathcal{K}_{0} + 6\mathcal{K}_1 - 3\mathcal{K}_2 \\
S^{\rm BB}_2 &= \mathcal{K}_{0} + 3\mathcal{K}_3 - 3\mathcal{K}_4 \\
S^{\rm BB}_3 &= \mathcal{K}_2 + 2\mathcal{K}_3 - 2\mathcal{K}_5 \\
S^{\rm BB}_4 &= 2\mathcal{K}_3 - \mathcal{K}_6,
\end{align}
where $S^{\rm BB}_1$ is the equilateral template. 

In our expansion of the non-Gaussian field, the number of operators that can generate a term in the bispectrum associated with $\{\mathcal{K}_n\}$ \cite{Byun:2013jba} is equal to the number of distinct entries in $\{p,r,s\}$. This redundancy allows us to cancel loop divergences in the first non-Gaussian correction to the power spectrum. The relationship between $\{\mathcal{K}_n\}$, all possible terms in $\Phi_2(x)$ that can generate that shape in the bispectrum, and the coefficients $\{a_i\}$ of the set of terms used in the body of the paper is shown in Table \ref{table:Kn_operators}. Some terms in $\Phi_2(x)$ are truly redundant for our purposes here, and so we did not include them (they are marked with an $\times$ in the last column of the table).

After imposing loop constraints, we imposed that the bispectrum is no more divergent than $k_l^{-1}$ in the squeezed limit and found four linearly independent solutions (Eq.\ref{kernels}) that generate shapes 
\begin{widetext}
\begin{align}
\label{templates_new}
&S_{1}(k_1,k_2,k_3) = \frac{1}{2} \bigg(\frac{k_1}{k_2} + cyc\bigg) - \frac{1}{2} \bigg(\frac{k_1^2}{k_2k_3} + cyc\bigg) - 1 \\
&S_{2}(k_1,k_2,k_3) = \frac{1}{4}\bigg(\frac{k_3^2}{k_1k_2} + cyc\bigg) - \frac{1}{4}\bigg(\frac{k_1^3}{k_3^2k_2}+ cyc \bigg) 
 + \frac{1}{2}\bigg(\frac{k_1k_2}{k_3^2}+cyc \bigg) \\
& S_{3}(k_1,k_2,k_3) = -\bigg(\frac{k_1^2}{k_2 k_3} + cyc\bigg) + \frac{1}{2}\bigg(\frac{k_1}{k_2} + cyc\bigg) + \frac{1}{2}\bigg(\frac{k_1^3}{k_3^2 k_2} + cyc\bigg) - \frac{1}{2}\bigg(\frac{k_2^2}{k_3^2} + cyc\bigg) \\
& S_{4}(k_1,k_2,k_3) = \bigg(\frac{k_1^2}{k_2 k_3} + cyc\bigg) -\bigg(\frac{k_1^3}{k_3^2 k_2} + cyc\bigg) + \bigg(\frac{k_1^4}{k_2^2 k_3^2} + cyc\bigg)\;.
\end{align} 
\end{widetext}
A little bit of algebra shows that these shapes are related to those of Byun and Bean by 

\begin{align}
S^{\rm BB}_1 &= 2 S_1, \\
S^{\rm BB}_2 &= -S_1 + S_3 + 2 S_2 \\ 
S^{\rm BB}_3 &= \frac{4}{3} S_2, \\
S^{\rm BB}_4 &= \frac{4}{3} S_2 - \frac{1}{3} S_4 \;.
\end{align}

Although it is maybe not immediately obvious that the procedure for constructing non-Gaussian fields $\Phi_{\rm NG}(x)$ with $1/k_l$ squeezed limit bispectra would lead to the same results as constructing bispectral shapes only, studying the structure of the loop constraints shows that the redundancy between $\mathcal{K}_n$ and $\Phi_2(x)$ is exactly enough to be sure the loop constraints are always satisfied. Notice that this is not the case for shapes that correspond to bispectra with $1/k_l^2$ divergence: $\mathcal{K}_0$ corresponds to only one field term, which does not have good loop behavior. So, although $\mathcal{K}_0$ appears to be a fine basis element at the level of bispectral shapes, we should not take $\partial^{-2}(\partial \phi)^2$ as a basis element of the quadratic term for a well-behaved non-Gaussian field.

\section{Checking the flat-only shape against linear combinations of operators}
\label{app:LC}
For the simplest single-clock models, it has been established that the equilateral and orthogonal templates are sufficient to cover essentially all of the parameter space of Lagrangians with the two cubic operators shown in Eq.(\ref{eff_action}), at small $c_s$ and with $\tilde{c}_s$ an order one number. Since the flat-only shape has very little overlap with either of those templates {\it and} they have very little overlap with each other, it is reasonable that the flat-only shape will have small overlap with any bispectrum generated by the simplest single-clock Lagrangian. 

However, as we compare the flat-only shape to various other operators in the more general single-clock Lagrangian, it is useful to keep in mind that small overlap with individual operators does not automatically imply small overlap with all linear combinations of operators. For example, the orthogonal template has small overlap with the bispectra from both the $\dot{\pi}^3$ and $\dot\pi(\partial_i\pi)^2 $ operators (cosines of $-0.31$ and $0.09$, respectively). But, because the bispectra from those operators have high overlap, there is a range of linear combinations of the two where the part of the shapes that are the same effectively cancel, leaving a shape with very high overlap with the orthogonal template. 

Although we have not made an exhaustive check of the flat-only template against all possible linear combinations of operators, it is straightforward and not too computationally intensive to compare at least a subset of possible linear combinations by brute force. For example, consider a test shape, $S_t$, that is to be compared to linear combinations from a list of reference shapes, $S^{(i)}_{\rm ref}$. Take two shapes from this basis, $S^{(1)}_{\rm ref}$ and $S^{(2)}_{\rm ref}$ and consider the shape made by a linear combination
\be
S_{\rm LC}=aS^{(1)}_{\rm ref}+bS^{(2)}_{\rm ref}\;.
\ee
If both the test shape and all reference shapes can be normalized in the same way (say, to 6 in the equilateral limit as is often done), then $a+b=1$. The cosine of the test shape and the linear combination is then a function of just one free parameter:
\begin{align}
{\rm Cos}(S_t, S_{\rm LC})=\frac{aS_t\cdot S^{(1)}_{\rm ref}+(1-a)S_t\cdot S^{(2)}_{\rm ref}}{|S_t||S_{\rm LC}|}
\label{eq:lincomb}
\end{align} 
where $|S_t|=\sqrt{S_t\cdot S_t}$. Notice that $a=0,1$ correspond to the cosines of the test shape with the individual reference shapes, and the linear combination asymptotes to $S_{\rm LC}\propto S^{(1)}_{\rm ref}- S^{(2)}_{\rm ref}$ for large $\pm a$. So, only a relatively small range of $a$ (eg, $-10\lesssim a\lesssim 10$) needs to be examined. 

For example, taking the orthogonal template as the test shape and the bispectra from operators $\dot{\pi}^3$ and $\dot\pi(\partial_i\pi)^2$ for the reference shapes, this technique uncovers the set of linear combinations that have high overlap with the orthogonal template. If the cosine of the test shape with the all possible pairs of the reference shapes does not show any significant increase over the cosines of the test shape with the individual reference shapes, it is unlikely that any more complicated linear combinations will lead to a high cosine. If a linear combination does result in a significantly higher cosine, one can add that linear combination to the set of reference shapes and again check cosines between the test shape and pairwise linear combinations of the reference shapes. We report in the text the various linear combinations of shapes from operators that we have checked using this method. Moreover, we generalize Eq.(\ref{eq:lincomb}) to test the overlapping degrees between the flat-only shape and the linear combination of standard set of local, orthogonal and equilateral templates and the pairs of this set, which do not exceed $\sim 0.6$. This implies that the flat-only shape is not highly constrained by these three shapes.

\pagebreak

\end{document}